\documentclass[%
reprint,
superscriptaddress,
 amsmath,amssymb,
 aps,
 prd,
]{revtex4-2}
\usepackage[dvipsnames]{xcolor}
\usepackage{graphicx}% Include figure files
\usepackage{dcolumn}% Align table columns on decimal point
\usepackage{bm}% bold math
\usepackage{subfig}
\usepackage{tabularx}
\usepackage{lineno}
\usepackage{hyperref}
\usepackage{amsfonts}
\usepackage{float}
\usepackage{url}
\usepackage{changes}

\newcommand{\Real}{\mathfrak{Re}}

\hypersetup{
colorlinks=true, % false: boxed links; true: colored links
linkcolor=Blue, % color of internal links
citecolor=Blue, % color of links to bibliography
filecolor=black, % color of file links
urlcolor=Blue % color of external links
}
\makeatletter
\begin{document}

\preprint{APS/123-QED}

\title{Response of a Pulsar Timing Array to 
an Internal Gravitational-wave Source}% Force line breaks with \\

\author{Houyuan Qi}
\affiliation{Department of Astronomy, School of Physics, Peking University, 100871 Beijing, China}

\author{Xian Chen}
\email{Contacting author: xian.chen@pku.edu.cn}
\affiliation{Department of Astronomy, School of Physics, Peking University, 100871 Beijing, China}
\affiliation{Kavli Institute for Astronomy and Astrophysics at Peking University, 100871 Beijing, China}

\author{Lin Wang}
\affiliation{State Key Laboratory of Radio Astronomy and Technology, Shanghai Astronomical Observatory, CAS, 80 Nandan Road, Shanghai 200030, P. R. China}

\author{Kuo Liu}
\affiliation{State Key Laboratory of Radio Astronomy and Technology, Shanghai Astronomical Observatory, CAS, 80 Nandan Road, Shanghai 200030, P. R. China}

\date{\today}

\begin{abstract}
Millisecond pulsars (MSPs) are abundant in globular clusters (GCs) and probably
also in galactic nuclei. They offer the potential to form a miniature pulsar
timing array (mini-PTA) to detect nanohertz gravitational-wave (GW) sources
located inside the array. Since the size of such an array is comparable to the
wavelength of GW, the conventional plane-wave approximation becomes invalid,
and near-field effects, including wavefront curvature, non-radiative self-field
of the GW source, and direct perturbation of pulsar by GW, become significant.
In this work, we incorporate these effects in a comprehensive model to
calculate the timing residual induced by a GW source inside a mini-PTA. We also
consider realistic GW source configurations in GCs (M15 and $\omega$ Centauri)
and in galactic nuclei (Sgr A* and M31), and find that for MSPs located
sufficiently close to the GW source (within a few wavelengths), the residual
can reach $1~\mu\mathrm{s}$ in GCs and up to milliseconds in galactic centers,
within the potential detection reach of current radio telescopes. Crucially,
when the pulsar lies within a few GW wavelengths of the source, the
non-radiative field dominates and causes the residual to rise much more steeply
(between $1/r_e^2$ and $1/r_e^4$, where $r_e$ is the distance to the source)
than the conventional far-field scaling ($1/r_e$). These results demonstrate
that mini-PTAs in GCs or galactic nuclei can serve as powerful probes of
otherwise invisible GW sources, including intermediate-mass and supermassive
black hole binaries.  
\end{abstract}

\maketitle

\section{Introduction}\label{sec:intro}

Precise timing of radio pulsars has long been proposed as a promising method of
detecting gravitational waves (GWs) \cite{Sazhin_1978,Detweiler_1979}.
Millisecond pulsars (MSPs), characterized by their short ($<30$ ms) and highly
stable spin periods, are particularly suitable for use in pulsar timing.  To
distinguish genuine GW signals from noise, it is necessary to monitor multiple
pulsars (i.e., a pulsar timing array, or PTA) and analyze the correlations in
their timing residuals \cite{1983ApJ...265L..39H,2010CQGra..27h4013H}.  Decades
of dedicated timing campaigns have improved the precision of timing residual
measurements to the nanosecond (ns) level, which has recently led to the detection of
a potential nanohertz GW background
\cite{2023ApJ...951L...8A,2023A&A...678A..50E,2023ApJ...951L...6R,2023RAA....23g5024X,2025MNRAS.536.1489M}.

Pulsar discoveries are particularly efficient in globular clusters (GCs). Since
the first detection of a pulsar in a GC \cite{1987Natur.328..399L}, over $330$
pulsars have been observed in these dense stellar systems \footnote{Pulsars in
GCs: \url{https://www3.mpifr-bonn.mpg.de/staff/pfreire/GCpsr.html}}.
Remarkably, more than $95\%$ of these pulsars are MSPs, reflecting rich
dynamical processes in GCs to spin up neutron stars. In fact, N-body
simulations predict that each GC may host $10-100$ MSPs
\cite{2019ApJ...877..122Y}. Correspondingly, future more sensitive telescopes,
such as the Square Kilometre Array (SKA), are expected to increase the total
number of GC pulsars to about $1700$ \cite{2025arXiv251216154B}.

The abundance of MSPs in GCs has motivated the idea of using their timing
residuals to search for compact binary GW sources within the same clusters
\cite{1993Ap&SS.208...93S}. This approach holds particular promise for probing
intermediate-mass black holes (IMBHs, $10^3-10^5M_\odot$). They are
theoretically predicted to form in GCs, and may form binaries with stellar-mass
black holes ($10M_\odot$) or other IMBHs (see early studies such as
\cite{2004ApJ...613.1143B,2006ApJ...642..427B,2006ApJ...640L..39G,2006ApJ...653L..53A}
and more recent works including
\cite{2013A&A...557A.135K,2014MNRAS.444...29L,2015MNRAS.454.3150G,2016ApJ...832..192H,2016ApJ...819...70M,2020ApJ...899..149R,2021A&A...652A..54A,2025ApJ...988...15L,2025MNRAS.tmp..434S}).

However, one theoretical problem arises when an MSP is located near a GW source.
The wavefront curvature becomes significant
\cite{1993Ap&SS.208...93S,1994PhRvD..50.3795F}, invalidating the plane-wave
approximation. To overcome this issue, several works have extended the theory
into the ``near-field'' regime, where the minimum distance between the GW
source and the line of sight to the pulsar (this distance is called ``impact
parameter'', denoted by $b$) can be as small as a few wavelengths of the GW. It
was found that in the case of small $b$, the leading-order effect on the
propagation of pulsar signals decays rapidly as $1/b^3$
\cite{1998PhRvD..58d4003D,1999PhRvD..59h4023K}. Later studies recognized that
this rapid decay results from assuming both the pulsar and the observer to be
infinitely far from the GW source. If the pulsar is placed near the wave zone
boundary, the signal can be substantially stronger \cite{2013MNRAS.430..305H}.
This ``edge effect'' is relevant in realistic GC configurations where the GW
wavelength could be comparable to the size of the cluster.  Indeed, in this
case the timing residuals can reach $500$ ns under favorable conditions
\cite{lommen05,jenet05}.

Beyond the near-field challenge, the full potential of PTAs is realized when
multiple MSPs are timed simultaneously. For GC MSPs, it has been suggested that
combining timing data from several MSPs within the same cluster can enhance
sensitivity to an isotropic GW background \cite{2021EPJP..136.1087M} or to GW
bursts originating within the cluster \cite{2017PhRvD..96l3016M}. Recently, a
detailed near-field calculation incorporating the edge effect identified
promising conditions for detecting IMBH binaries in clusters such as M15 and
$\omega$ Cen, where multiple MSPs have been found near the cluster center and
could collectively achieve a timing-residual sensitivity at the microsecond
($\mu$s) level \cite{chen25}.

The idea and methodology have also been applied to MSPs in galactic centers.
Earlier proposals suggested that timing MSPs near the
Galactic Center could reveal a binary companion of Sgr A$^*$
\cite{2004ApJ...615..253P,2012ApJ...752...67K}.  The observational consequence
of a curved wavefront was also studied
\cite{2022ApJ...939...55G,2023ApJ...946...76K}, though initially in a different
context \cite{2021PhRvD.104f3015D}, and these studies suggest that monitoring a
sample of MSPs within about $1$ pc of Sgr A$^*$ could yield high
signal-to-noise detections of a potential binary companion. Similar analyses
have been extended to MSPs in nearby galaxies such as the Large Magellanic
Cloud, M31, M32, and M87, suggesting that future observations with the SKA
could detect binary GW sources if MSPs within $0.1-1$ pc of the centers are
monitored over a decade \cite{2025ApJ...978..104G}. 

In this work, we take a further step toward a comprehensive model of the PTA
response to an internal GW source by incorporating four elements essential for
realistic astrophysical configurations.
(i) Binary eccentricity: existing models
typically assume circular binaries as the GW sources, whereas dynamical
simulations of star clusters favor eccentric orbits. 
(ii) Second-order stress-energy tensor:  previous calculations (including ours \cite{chen25})
sometimes retain only the lowest-order terms in the stress-energy tensor, but
higher-order contributions may be equally important (e.g., see
Ref.~\cite{jenet05}).
(iii) Non-radiative self-field: close to the GW source
the near-field non-wave part of the strain induced by varying gravitational potential
becomes non-negligible but has sometimes been overlooked in previous works.
(iv) Pulsar
term: the effect of GWs on the pulsar itself is often neglected but becomes
important when the pulsar lies within a few GW wavelengths of the source.

The paper is organized as follows. Section~\ref{sec:near} presents our theoretical
framework, including the derivation of the timing residual in the near-field
regime, the expansion of the GW strain to second order, and the calculation of
the mass quadrupole moment for eccentric binaries. In Section~\ref{sec:results}, we analyze
the near-field effects and compute the timing residuals for specific
astrophysical systems including the GCs M15 and $\omega$ Centauri, as well
as the galactic nuclei Sgr A$^*$ and M31. We also examine the phase differences
between MSPs and discuss the detectability of the signal. Section~\ref{sec:dis} 
discusses our results, compares our model predictions with previous works, and presents
our conclusions.
The code and the numerical data used to produce the figures are available at \footnote{\url{https://github.com/HouyuanQi/PTA-Near_field_timing_residual}}.

\section{Near-field Timing Residual}\label{sec:near}

\subsection{Timing Residual}

\begin{figure}
    \centering
    \includegraphics[width=1\linewidth]{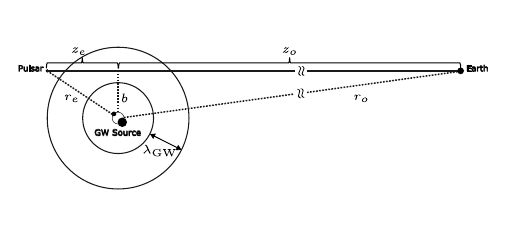}
    \caption{System configuration and definition of the parameters. Also see Ref.~\cite{jenet05}.}
    \label{fig:configuration}
\end{figure}

Fig.~\ref{fig:configuration} illustrates the system of our interest.  It
includes a black-hole binary as the GW source, an MSP (emitter), and Earth (observer).
The MSP and the GW source are located in a GC far from the observer with
$r_e\ll r_o$.  We choose a coordinate system such that the center-of-mass of
the black-hole binary is static and located at the origin. We further assume that the
MSP and the Earth are also static, and their locations are $(b,0,z_e)$ and
$(b,0,z_o)$, respectively.  The pulses emitted by the MSP propagate along the
$z$ axis with an impact parameter of $b$.  If we consider nanohertz GWs, the
wavelength is approximately $\lambda_{\mathrm{GW}}\sim\mathrm{pc}$, comparable
to the size of a typical GC. Consequently, both the MSP and the pulse signals it
emits reside in the near-field zone, where the non-radiative component of the
gravitational field is significant.

The passing GW perturbs the spacetime between the pulsar and the Earth, thereby causing the observed pulse arrival times to deviate from those in flat spacetime.
The time difference, also known as the timing residual $R(t)$, is given by 
\begin{align}
    R(t)=\int_{0}^{t}\frac{\nu_0-\nu(t_{\rm obs})}{\nu_{0}}\,\mathrm{d} t_{\rm obs},
\end{align}
where $\nu_0=\left.-g^{\mu\nu}K_{\mu}V_{\nu}\right|_{e}$ is the emitted
frequency in the rest frame of the pulsar, and
$\nu=\left.-g^{\mu\nu}K_{\mu}V_{\nu}\right|_{o}$ is the frequency measured at
the observer at the time $t_{\rm obs}$.  Here,
$g^{\mu\nu}=\eta^{\mu\nu}-h^{\mu\nu}$ denotes the perturbed metric of
spacetime, $K_{\mu}$ is the dual to the photon's four-momentum, and $V_{\mu}$
is the dual to the four velocity, where $e$ and $o$ denote the emitter (pulsar) and the
Earth, respectively.

In the weak-field limit we have $h^{\mu\nu}\ll1$, and the four-vectors take
the form $K_{\mu}=\bar{K}_{\mu}+\delta K_{\mu}$, $V_{\mu}=\bar{V}_{\mu}+\delta
V_{\mu}$. The barred vectors $\bar{K}_{\mu}=\nu_0/c\,(-1,0,0,1)$ and
$\bar{V}_{\mu}=c\,(-1,0,0,0)$ represent unperturbed values in flat spacetime,
satisfying $-\eta^{\mu\nu}\bar{K}_{\mu}\bar{V}_{\nu}=\nu_0$.

With the above definitions, we find that in leading order,
\begin{align}
    \frac{\nu_0-\nu}{\nu_0}=&\left.\frac{-h_{\mu\nu}\bar{K}^{\mu}\bar{V}^{\nu}+\bar{K}^{\mu}\delta V_{\mu}+\bar{V}^{\mu}\delta K_{\mu}}{\nu_0}\right|_{o}\nonumber\\
    =&\left.\left[-h_{tt}-h_{tz}+\frac{1}{c}(\delta V_{t}+\delta V_{z})+\frac{c}{\nu_0}\delta K_{t}\right]\right|_{o},
\end{align}
where the term $-h_{\mu\nu}\bar{K}^{\mu}\bar{V}^{\nu}$ was neglected in our previous work \cite{chen25}.
For $\delta K_{\mu}$ and $\delta V_{\mu}$, they are governed by geodesic equations
\begin{align}
    \frac{d\,\delta K_\alpha}{d\lambda}=&\frac{1}{2}h_{\mu\nu,\alpha}\bar{K}^{\mu}\bar{K}^{\nu},\\
    \frac{d\,\delta V_{\alpha}}{d\tau}=&\frac{1}{2}h_{\mu\nu,\alpha}\bar{V}^{\mu}\bar{V}^{\nu}.
\end{align}
Neglecting the motion of the pulsar and the observer, we have
\begin{align}
	\left.\delta K_{t}\,\right|_o=&\left.\delta K_t\,\right|_e+\frac{\nu_0}{c}\int_{z_e}^{z_o}\frac{h_{tt,t}+h_{zz,t}+2h_{tz,t}}{2}\,\mathrm{d}z,\label{eq:delta K}\\
    \delta V_{t}=&c^2\int_0^t\frac{h_{tt,t}}{2}\,\mathrm{d}t,\qquad\delta V_{z}=c^2\int_0^t\frac{h_{tt,z}}{2}\,\mathrm{dt}.\label{eq:delta V}
\end{align}
In Equation~(\ref{eq:delta K}),
$\left.\delta K_{t}\right|_{e}$ arises from the GW perturbation of a photon just emitted from the pulsar, and
can be solved from $-\left.g^{\mu\nu}(\bar{K}_{\mu}+\delta K_{\mu})(\bar{V}_{\nu}+\delta V_{\nu})\right|_e=\nu_0$. To linear order, it is
\begin{align}
    \left.\delta K_{t}\right|_{e}=\left.\frac{\nu_0}{c}(h_{tt}+h_{tz})-\frac{\nu_0}{c^2}(\delta V_t+\delta V_z)\right|_{e}.
\end{align}
The two perturbations shown in Equation~(\ref{eq:delta V}) are applicable 
not only to the four velocity of the observer but also to that of the pulsar.

Finally, the frequency difference can be expressed using $h_{\mu\nu}$ as
\begin{align}
    \frac{\nu_0-\nu}{\nu_0}=&\int_{z_e}^{z_o}\frac{h_{tt,t}+h_{zz,t}+2h_{tz,t}}{2}\,\mathrm{d}z\nonumber\\
    &+\left.\left[-h_{tt}-h_{tz}+c\int\frac{h_{tt,t}+h_{tt,z}}{2}\,\mathrm{d}t\right]\right|^{o}_{e},\label{eq:Rt}
\end{align}
where the first integration is along the photon's geodesic, and the second one is along the observer's and pulsar's geodesics. The bracket evaluated at the observer is called the Earth term, and that evaluated at the pulsar is called the pulsar term. For the photon, we choose $z$ and the observed time $t_{\rm obs}$ as parameters, so the local time $t$ is given by $t=t_{\rm obs}-(z_o-z)/c$. The upper limits of integration in the Earth term and in the pulsar term are $t_{\rm obs}$ and $t_{\rm obs}-(z_o-z_e)/c$.

So far,
we have shown that the complete form of the timing residual contains both the
Earth and the pulsar terms. However, in the context studied in this paper,
where the pulsar is relatively close to the GW source and the Earth is far from
the source and the pulsar (typically satisfying $r_o/r_e\gtrsim10^5$), the
Earth term can be neglected, so we omit it in the later calculations.

\subsection{GW Strain to the Second Order}

Now we evaluate the GW strain $h_{\mu\nu}$ 
at a position $\mathbf{r}$ and the local time $t$. 
We start from the linearized Einstein equation
\begin{equation}
    \square \bar{h}_{\mu\nu}=-\frac{16\pi G}{c^4}T_{\mu\nu},
\end{equation}
where $T_{\mu\nu}$ is the stress-energy tensor of the GW source and $\bar{h}_{\mu\nu}=h_{\mu\nu}-\frac{1}{2}\eta_{\mu\nu}h$ is the trace-reversed GW strain. The retarded solution at $(t,\mathbf{r})$ is given by
\begin{equation}
	\bar{h}_{\mu\nu}(t,\mathbf{r})=\frac{4G}{c^4}\int\frac{T_{\mu\nu}(t-|\mathbf{r}-\mathbf{r}'|/c,\mathbf{r}')}{|\mathbf{r}-\mathbf{r}'|}\,\mathrm{d}^3\mathbf{r}',\label{eq:hretarded}
\end{equation}
where $\mathbf{r'}$ refers to the source position, with $|\mathbf{r'}|$ comparable to the size of the GW source.
Notice that the solution is not in the transverse-traceless gauge.

In this work, the GW source is an IMBH binary on bounded Keplerian motion with
a semimajor axis of $a$.  Moreover, we are interested in the case where
$|\mathbf{r}|\ge b
\gg |\mathbf{r}'| \sim a$, where $b$ is the impact parameter defined in Section~\ref{sec:intro}.  
Under these conditions, we can Taylor expand the integrand in
Equation~(\ref{eq:hretarded}) in different orders of $\mathbf{r'}$, which gives
\begin{widetext}
    \begin{align}  
    &\bar{h}_{\mu\nu}\sim \frac{4G}{c^4}\int\left(\frac{1}{r}+\frac{x_ix_i'}{r^3}-\frac{1}{2}\frac{x_i'x_i'}{r^3}+\frac{3}{2}\frac{x_i'x_j'x_ix_j}{r^5}\right)  
    \times\left[1+\left(\frac{x_ix_i'}{r}-\frac{1}{2}\frac{x_i'x_i'}{r}+\frac{x_i'x_j'x_ix_j}{2r^3}\right)\frac{\partial_t}{c}+\frac{1}{2}\frac{x_i'x_j'x_ix_j}{r^2}\frac{\partial_t^2}{c^2}\right]T_{\mu\nu}\,\mathrm{d}^3\mathbf{r}'.
    \end{align} 
\end{widetext}
Here, we have adopted the convention that the Latin letters, such as $i$ and $j$, refer only to the three spatial components. 

To understand the magnitudes of the time derivatives in the above equation, 
one can use $\partial_tT\sim\omega T$, where $\omega$ is the average angular frequency of the binary. Since 
\begin{align}
    \frac{x'}{r}\sim\frac{a}{b}\ll 1,\quad \frac{x'\partial_t}{c}\sim\frac{a\omega}{c}\sim\frac{v}{c}\ll 1,
\end{align}
where $v$ is the typical orbital velocity of the binary, the $(n+1)$-th term
from the Taylor expansion is smaller than the  $n$-th term by a factor of
either $x'/r$ or $v/c$. We find that in our near-field cases the two factors
$v/c$ and $a/b$, although small, can be comparable to each other, so we keep
both the $v/c$ and the $a/b$ terms in our calculation.

It is important to note that
besides the terms at different Taylor orders, the components of
$\bar{h}_{\mu\nu}$ are also of different orders of $v/c$. 
More specifically, 
$\bar{h}_{\mu\nu}$ is proportional to $T_{\mu\nu}$ according to Einstein's equation, and
the leading terms of $T_{\mu\nu}$ in Newtonian approximation are 
\begin{align}
    T_{\mu\nu}\sim\rho
    \begin{pmatrix}
    c^2&cv_i\\
    cv_j&v_iv_j
    \end{pmatrix}.
\end{align}
Therefore, $\bar{h}_{ij}\sim (v/c)\bar{h}_{ti}\sim(v/c)^2\bar{h}_{tt}$, i.e.,
the $0$th order term (in the Taylor expansion) of $\bar{h}_{ij}$ is comparable
to the first order term of $\bar{h}_{ti}$ and the second order term of
$\bar{h}_{tt}$. For this reason and to be complete up to the second order of
$v/c$ or $a/b$, in the following we retain $\bar{h}_{ij}$ up to the leading
order, $\bar{h}_{ti}$ up to the next order, and $\bar{h}_{tt}$ up to the second
order after the leading one. This model is different from our previous one
\cite{chen25} in which all the $\bar{h}_{\mu\nu}$ components only contain the
leading order terms, and more similar to the original
model proposed in Ref.~\cite{jenet05}. We will show later that this difference will
significantly affect the results. 

Including all the terms up to the second order, we find that the integration gives
\begin{eqnarray}
    \bar{h}_{tt}&=&\frac{4GM}{rc^2}+\frac{2G}{r^3c^2}\frac{3x_ix_j-r^2\delta_{ij}}{r^2}Q_{ij}\nonumber\\
    &+&\frac{2G}{r^2c^3}\frac{3x_ix_j-r^2\delta_{ij}}{r^2}\dot{Q}_{ij}+\frac{2G}{rc^4}\frac{x_ix_j}{r^2}\ddot{Q}_{ij},\label{eq:htt}\\
    \bar{h}_{ti}&=&-\frac{2G}{r^2c^3}\epsilon_{ijk}\frac{x_j}{r}J_k-\frac{2G}{r^2c^3}\frac{x_j}{r}\dot{Q}_{ij}-\frac{2G}{rc^4}\frac{x_j}{r}\ddot{Q}_{ij},\label{eq:hti}\\
    \bar{h}_{ij}&=&\frac{2G}{rc^4}\ddot{Q}_{ij},\label{eq:hij}
\end{eqnarray}
where $M=\int (T_{00}/c^2)d^3\mathbf{r}'=m_1+m_2$ is the total mass of the
binary GW source, $Q_{ij}=Q^{ij}=\int T_{00}x'^i x'^j/c^2d^3\mathbf{r}'$ is the
mass quadrupole moment of the source, and we have set the total momentum
$P_i=0$ and the center of mass $x_{c,i}=0$, and assumed that the total angular momentum $J_i$ is conserved. Using these expressions, one can verify that
$\partial^\mu\bar{h}_{\mu\nu}=0$, which agrees with the continuity equation of
$T_{\mu\nu}$. Another useful finding is that only the
second-order terms of $\bar{h}_{tt}$ and the first-order terms of
$\bar{h}_{tz}$ are time dependent, so that even though the lower-order terms 
are larger, they do not affect the timing residual. 

\subsection{Calculating the Mass Quadrupole}

Now we derive the expressions for the quadrupole terms, bearing in mind that
the binary orbit can be eccentric. Given a binary with a reduced mass of 
$\mu=m_1m_2/(m_1+m_2)$, a semimajor axis of $a$, and an eccentricity of $e$, 
the quadrupoles are given by
\begin{equation}
    Q_{ij}=\mu x_ix_j,
\end{equation}
where $\mathbf{x}=\mathbf{x}_1-\mathbf{x}_2$ is the relative position of the
two members.  The orientation can be described using Euler angles, with
$\theta,\phi,\psi$ representing the nutation angle, the precession angle, and
the rotation angle of the perihelion, respectively. We further assume that the orbital plane
of the binary is the $x'-y'$ plane, the center of mass coincides with the origin,
and the perihelion is along the positive $x'$-axis. Then the coordinates 
$x_i'(t)$ ($i=1,2,3$) in the orbital plane
are related to the coordinates $x_i(t)$ in the source frame by a rotation,
\begin{align}
    \mathbf{x}=R_{1}(\phi)R_{2}(\theta)R_{3}(\psi)\mathbf{x'}\equiv R_{\rm rot}\mathbf{x'},
\end{align}
where $R_i$ is the rotation matrix corresponding to a Euler angle. 
With this coordinate transformation, $Q_{ij}$ can be expressed as
\begin{align}
    Q_{ij}=\mu (R_{\rm rot}\mathbf{x'}\mathbf{x'}^{T}R_{\rm rot}^{T})_{ij}.
\end{align}

For an elliptical Keplerian motion, the components of $\mathbf{x'}\mathbf{x'}^{T}$ can be decomposed into Fourier series:
\begin{align}
    \mathbf{x'}\mathbf{x'}^{T}=\Real(\sum_{n=0}^\infty a^2\tilde{A}_n\mathrm{e}^{\mathrm{i}n\omega t}),
\end{align}
where $\omega=\sqrt{GM/a^3}$ is the average angular frequency of the motion. If the orbit is circular, there are only $n=0,2$ components. In more general cases, we have
\begin{align}
    \tilde{Q}_{ij}=\mu a^2\sum_{n=0}^\infty\tilde{q}_{ij,n}\mathrm{e}^{\mathrm{i}n\omega t},\label{eq:Qij}
\end{align}
where $\tilde{q}_{ij,n}=(R_{\rm rot}\tilde{A}_nR_{\rm rot}^{T})_{ij}$ is the factor determined by
orientation and eccentricity, and we have introduced the complex form of
$\tilde{Q}_{ij}$ for simplicity, which satisfies
$\Real(\tilde{Q}_{ij})=Q_{ij}$. The explicit form of $\tilde{A}_n$ is provided in
Appendix~\ref{app:qijn}.

\subsection{Expression of $R(t)$}

Having specified the system of our interest, now we can calculate 
the timing residual  $R(t)$ to the second order of $v/c$ or $a/b$. The
calculation is carried out in a complex representation, and the physical
results are given by the corresponding real parts. Since we
are interested in the oscillating part of the timing residual, terms that do
not vary with time will be neglected. Choosing $x=b$, $y=0$, and using
Equations~(\ref{eq:htt}), (\ref{eq:hti}), (\ref{eq:hij}) and (\ref{eq:Qij}), we
can rewrite the GW strains appearing in Equation~(\ref{eq:Rt}) as
\begin{align}
    h_{tt}+h_{zz}&+2h_{tz}=\frac{2\mu a^2G}{r^3c^2}\sum_{n=1}^\infty\tilde{\xi}_n\mathrm{e}^{\mathrm{i}n\omega t_r}\label{eq:q1},\\
    h_{tt}&=\frac{\mu a^2 G}{r^3c^2}\sum_{n=1}^\infty\tilde{\zeta}_n\mathrm{e}^{\mathrm{i}n\omega t_r},\label{eq:q2}\\
    h_{tz}&=\frac{\mu a^2G}{r^3c^2}\sum_{n=1}^{\infty}\tilde{\eta}_{n}\mathrm{e}^{\mathrm{i}n\omega t_r},
\end{align}
where we have defined the dimensionless functions 
\begin{widetext}
    \begin{align}
         \tilde{\xi}_n&=
    \left[\left(\frac{3b^2}{r^2}-1\right)\left(1+r\frac{\mathrm{i}\omega n}{c}\right)-b^2\frac{n^2\omega^2}{c^2}\right]\tilde{q}_{xx,n}
    +\left[\frac{3zb}{r^2}+b\left(\frac{3z}{r}-1\right)\frac{\mathrm{i}\omega n}{c}-b\left(\frac{z}{r}-1\right)r\frac{n^2\omega^2}{c^2}\right](\tilde{q}_{zx,n}+\tilde{q}_{xz,n})\nonumber\label{eq:xin}\\
    &-\left(1+r\frac{\mathrm{i}\omega n}{c}\right)\tilde{q}_{yy,n}
    +\left[\left(\frac{3z^2}{r^2}-1\right)+\left(\frac{3z^2}{r^2}-\frac{2z}{r}-1\right)r\frac{\mathrm{i}\omega n}{c}-\left(\frac{z}{r}-1\right)^2r^2\frac{n^2\omega^2}{c^2}\right]\tilde{q}_{zz,n},\\
         \tilde{\zeta}_n&=
    \left[\left(\frac{3b^2}{r^2}-1\right)\left(1+r\frac{\mathrm{i}\omega n}{c}\right)-\left(b^2+r^2\right)\frac{n^2\omega^2}{c^2}\right]\tilde{q}_{xx,n}
    +\left[\frac{3zb}{r^2}\left(1+r\frac{\mathrm{i}\omega n}{c}\right)-bz\frac{n^2\omega^2}{c^2}\right](\tilde{q}_{zx,n}+\tilde{q}_{xz,n})\nonumber\\
    &-\left(1+r\frac{\mathrm{i}\omega n}{c}+r^2\frac{n^2\omega^2}{c^2}\right)\tilde{q}_{yy,n}
    +\left[\left(\frac{3z^2}{r^2}-1\right)\left(1+r\frac{\mathrm{i}\omega n}{c}\right)-\left(z^2+r^2\right)\frac{n^2\omega^2}{c^2}\right]\tilde{q}_{zz,n},\label{eq:zetan}\\
        \tilde{\eta}_{n}&=
    \left(-2b\frac{\mathrm{i}\omega n}{c}+2br\frac{n^2 \omega^2}{c^2}\right)\tilde{q}_{zx,n}
    +\left(-2z\frac{\mathrm{i}\omega n}{c}+2zr\frac{n^2 \omega^2}{c^2}\right)\tilde{q}_{zz,n}\label{eq:etan}.
    \end{align}
\end{widetext}
From the expression of $\tilde{\xi}_n$ we find 
that it scales as $r$, so that $h_{tt}+h_{zz}+2h_{tz}\sim O(1/r^2)$. 
Therefore, significant contribution to the integral only arises at small $r$, in contrast to 
the conventional understanding of quadrupole terms which decay as $O(1/r)$. 
For this reason, the impact of GW on light is localized.

Now we calculate the two parts of the timing residual
in Equation~(\ref{eq:Rt}).
For the first part, using Equation~(\ref{eq:q1}) and integrating over time gives
\begin{align}
    \tilde{R}_1(t)&=
    \frac{\mu a^2G}{c^3}\sum_{n=1}^\infty\int_{z_e}^{z_o}\frac{\tilde{\xi}_n(z)}{r^3}\nonumber\\
    &\qquad\times\exp[\mathrm{i}\omega n(t-\frac{z_o-z}{c}-\frac{r}{c})]\,\mathrm{d}z.
\end{align}
For the second part, the integration operates along the pulsar's geodesic. 
Actually, the integrand can be written as
\begin{align}
    \frac{c}{2}(&h_{tt,t}-h_{tt,z}+2h_{tz,t})=\nonumber\\
    &\frac{\mu a^2G}{c}\sum_{n=1}^{\infty}\mathrm{e}^{\mathrm{i}n\omega t}
    \Biggl[\frac{1}{2}\biggl(\frac{\mathrm{i}n\omega}{c}-\partial_z\biggr)\Biggl(\frac{\tilde{\zeta}_n(z)}{r^3}\mathrm{e}^{-\mathrm{i}n\omega r/c}\Biggr)\nonumber\\
    &+\biggl(\frac{\mathrm{i}n\omega}{c}\frac{\tilde{\eta}_{n}(z)}{r^3}\mathrm{e}^{-\mathrm{i}n\omega r/c}\biggr)\Biggr].
\end{align}
Integrating twice over time, the oscillating part gives
\begin{align}
    \tilde{R}_{2}(t)=&-\frac{\mu a^2G}{c\omega^2}\sum_{n=1}^\infty\frac{1}{n^2}\mathrm{e}^{\mathrm{i}n\omega(t-(z_o-z_e)/c)}\nonumber\\
    &\times\Biggl[\frac{1}{2}\biggl(\partial_z-\frac{\mathrm{i}n\omega}{c}\biggr)\Biggl(\frac{\tilde{\zeta}_n(z)}{r^3}\mathrm{e}^{-\mathrm{i}n\omega r/c}\Biggr)\nonumber\\
    &-\biggl(\frac{\mathrm{i}n\omega}{c}\frac{\tilde{\eta}_{n}(z)}{r^3}\mathrm{e}^{-\mathrm{i}n\omega r/c}\biggr)\Biggr]\Biggr|^o_{e}.
\end{align}

To account for systems of different parameters, such as binary
mass, pulsar position, and orbital period ($P$), we scale length
($r$, $z$ and $b$) using the wavenumber $k=2\pi/\lambda_{GW}=2\pi/(cP)$. Then the
timing residuals can be expressed in dimensionless quantities such as $kr$, $kz$, and $kb$,
and the results are
\begin{align}
    \tilde{R}_1(t)&=\frac{G^2m_1m_2}{ac^5}\sum_{n=1}^{\infty\nonumber}\\
    &\qquad\int_{kz_e}^{kz_o}\,\mathrm{d}(kz)\,\frac{\tilde{\xi}_n(kz)}{(kr)^3}\mathrm{e}^{\mathrm{i}n(kz-kr-kz_o)}\mathrm{e}^{\mathrm{i}n\omega t}\nonumber\\
	&=\mathrm{Amp}\cdot\sum_{n=1}^{\infty}\int_{kz_o}^{kz_e}\tilde{r}_{1,n}\mathrm{e}^{\mathrm{i}n\omega t}\,\mathrm{d}(kz)\label{eq:r1n}\\&=\mathrm{Amp}\cdot\sum_{n=1}^{\infty}\tilde{R}_{1,n}\mathrm{e}^{\mathrm{i}n\omega t}\label{eq:R1t},
\end{align}
and
\begin{align}
	\tilde{R}_2(t)=&\frac{G^2m_1m_2}{ac^5}\sum_{n=1}^{\infty}\frac{1}{n^2}\mathrm{e}^{\mathrm{i}n\omega t}\mathrm{e}^{\mathrm{i}n(kz-kz_o)}\nonumber\\
	\times&\Biggl[\frac{1}{2}\biggl(\mathrm{i}n-\frac{\partial}{\partial(kz)}\biggr)\Biggl(\frac{\tilde{\zeta}_n(kz)}{(kr)^3}\mathrm{e}^{-\mathrm{i}n(kr)}\Biggr)\nonumber\\
    &+\biggl(\mathrm{i}n\frac{\tilde{\eta}_{n}(kz)}{(kr)^3}\mathrm{e}^{-\mathrm{i}n(kr)}\biggr)\Biggr]\Biggr|^o_e\\
    =&\mathrm{Amp}\cdot\sum_{n=1}^{\infty}\tilde{R}_{2,n}\mathrm{e}^{\mathrm{i}n\omega t}.\label{eq:R2t}
\end{align}
The scaling amplitudes are the same in both equations,
\begin{align}
	\mathrm{Amp}&=\frac{G^2m_1m_2}{ac^5}\\
	&\simeq\frac{0.23q~\mu{\rm s}}{(1+q)^2}\left(\frac{P}{1~{\rm yr}}\right)^{-2/3}
	\left(\frac{M}{10^4~M_\odot}\right)^{5/3}\\
	&\simeq\frac{1.1~{\rm ms}}{(1+q)^2}\left(\frac{q}{10^{-3}}\right)\left(\frac{P}{1~{\rm yr}}\right)^{-2/3}
	\left(\frac{M}{10^8~M_\odot}\right)^{5/3}\label{eq:Amp},
\end{align}
and it depends on the
orbital period $P$, the total mass $M$, and the mass ratio $q=m_2/m_1$ of the two compact objects
in the binary GW source. 
Notice that the dimensionless function $\tilde{R}_2(t)$ can be computed readily, and 
$\tilde{R}_1(t)$ also has an analytical expression. The analytical expressions of them are presented in Appendix~\ref{app:R1(t)}.
If the binary is circular, we
only need to consider the $\tilde{R}_{1,n=2}$ and $\tilde{R}_{2,n=2}$ terms.

\section{Results}\label{sec:results}

\subsection{Near-field Effects}

\begin{figure}
    \centering
    \includegraphics[width=\linewidth]{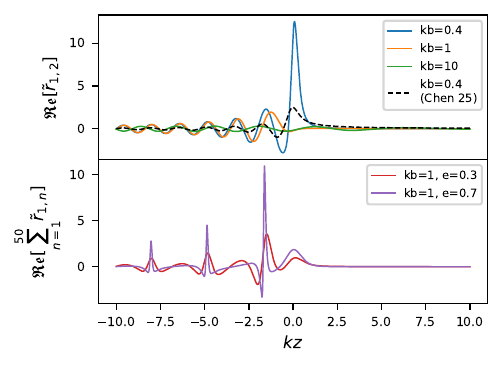}
    \caption{  
    The real part of the integrand $\sum\tilde{r}_{1,n}$ that appears in the first part of the timing residual.
    For the orientation of the binary, we set $\theta=\pi/4$ and $\phi=\psi=0$. 
    The upper panel shows $\Real(\tilde{r}_{1,2})$ for a circular binary, with different solid curves correspond to different impact parameters, and the dashed curve is taken from our previous work where we kept only the leading-order term of the stress-energy
	tensor \cite{chen25}.
    The lower panel shows the sum of $\Real(\tilde{r}_{1,n})$ for the first 50 harmonics for an elliptical orbit with different eccentricity.
    }
    \label{fig:Rer12}
\end{figure}

Because the pulsar in our model is close to the GW source, the wavefront is curved at the pulsar
and the non-radiative self-field of the GW source is also important. 
To highlight these ``near-field effects'', Fig.~\ref{fig:Rer12} shows
the real part of the integrand, $\tilde{r}_{1,n}$, which is used in the
calculation of $\tilde{R}_{1,n}$ (see Eq.~(\ref{eq:r1n})). 
In the upper panel, we assume that the binary undergoes circular orbit, so we only keep $n=2$ term, while in the lower panel we considered eccentric orbits.

The result depends on $kb$. When $kb$ is small (blue curve), the function at
the location of $|kz|\lesssim 1$ can significantly exceed unity, highlighting
the strong effect within a distance of one wavelength from the source.
Compared to our earlier calculation (dashed curve, see Ref.~\cite{chen25}) where
we kept only the leading term of the stress-energy tensor, the effect in the
current complete model is more prominent. 
For a more general $kb$ (orange or green curve), we see that the function decays as $O(1/r^2)$ for positive $z$, as we have discussed 
before. For negative $z$, the function $\tilde{r}_{1,2}$ is oscillating and the magnitude
decreases as $O(1/r)$. Because of the oscillation, the net contribution to timing residual is small.

The lower panel shows that for elliptical orbits, the integrand, after summing up the Fourier components, exhibit several sharp spikes. These spikes arise from the stronger GW emission at the binary's pericenter. And larger eccentricity leads to narrower and shaper spikes.

\begin{figure*}[htbp]
    \centering
    \begin{minipage}{0.48\linewidth}
        \centering
        \includegraphics[width=\linewidth]{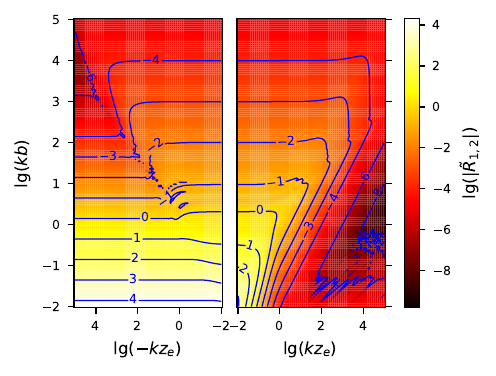}
    \end{minipage}\hfill
    \begin{minipage}{0.48\linewidth}
        \centering
        \includegraphics[width=\linewidth]{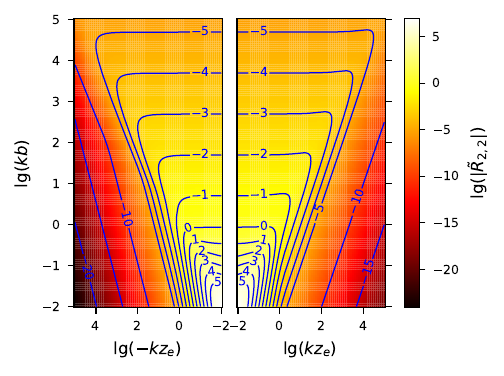}
    \end{minipage}
    \caption{Dimensionless timing residuals $|\tilde{R}_{1,n=2}|$ (left) and $|\tilde{R}_{2,n=2}|$ (right) as a function of the scaled position $kz_e$ and scaled impact parameter $kb$ of the pulsar. 
The orientation and eccentricity parameters are the same as in Fig.~\ref{fig:Rer12}.}
    \label{fig:R12_R22}
\end{figure*}

To see the net effect and its dependence on pulsar position, we show in the left panel 
 of Fig.~\ref{fig:R12_R22} the integrated dimensionless timing residual 
 $|\tilde{R}_{1,n=2}|$ as a function of $kz_e$ and $kb$. The observer
is put at infinity.
Notice that both positive and negative $kz_e$ are considered. When $kz_e$ is positive,
the pulsar lies between the GW source and the observer, and the residual is large if
both $kz_e$ and $kb$ are small. When $kz_e$ is negative, the pulsar lies
behind the source. In this case, as long as $kb$ is small, the emitted light will
pass closely by the GW source and produce a large timing residual (indicated by a
large area with bright yellow color).

The right panel of Fig.~\ref{fig:R12_R22} shows the  
dimensionless  pulsar term, $\tilde{R}_{2,n=2}$, for the same system configuration. 
The symmetry between positive and negative $kz_e$ reflects the dependence on the
distance $\sqrt{b^2+z_e^2}$. This dependence is also the reason that $|\tilde{R}_{2}|$  
is large when the pulsar is close to the source. In fact, this pulsar term 
can be comparable to $|\tilde{R}_1|$ even at large $kb$, e.g., $kb\sim10$,
if one compares the two panels in Fig.~\ref{fig:R12_R22}.
This result indicates    that the longer the period of the GW source is, the more important
the pulsar term becomes.

It is worth noting that
when the orbital period of the GW source is long (or $kb$ is small), our earlier assumption, that $a/b$ is sufficiently small,
may break down. To be self-consistent, we therefore require that
\begin{align}
    kb\gg0.0461\left(\frac{P}{1\,\mathrm{yr}}\right)^{-1/3}\left(\frac{M}{10^8\,M_{\odot}}\right)^{1/3}.
\end{align}
In this regime, our results are accurate.

Finally, Fig.~\ref{fig:R12_R22} and Eq.~(\ref{eq:Amp}) provide a method to estimate the timing residual of a specific system. 
One first calculates the amplitude factor from Eq.~(\ref{eq:Amp}) using the binary parameters, and then reads the dimensionless $|\tilde{R}_{1,2}+\tilde{R}_{2,2}|$ from Fig.~(\ref{fig:R12_R22}). 
Their product gives an order-of-magnitude estimate of $R(t)$. The phase difference between $\tilde{R}_{1,2}$ and $\tilde{R}_{2,2}$, as well as orientation effects, are not included, but they are less dominant.

\subsection{Timing Residuals for MSPs in Globular Clusters}

Now we apply the complete model presented above to MSPs in GCs.  We consider an
IMBH binary with a total mass of $m_1+m_2=(10^3-10^4)\,M_{\odot}$ and an
orbital period of $P=(0.1-10)\,\mathrm{yr}$. If the binary orbit is circular, 
the merger timescale due to GW radiation is
\begin{align}
    T_{\mathrm{GW}}&\simeq\frac{a}{4|\dot{a}|}=\frac{5c^5a^4}{256G^3m_1m_2(m_1+m_2)}\\
    \simeq&6.94\times10^{10}\,\mathrm{yr\,}\frac{(1+q)^2}{q}\left(\frac{M}{10^4\,M_{\odot}}\right)^{-5/3}\left(\frac{P}{1\,\mathrm{yr}}\right)^{8/3}
\end{align}
\cite{Peters_1963}.
Therefore, the binary evolves slowly and remains stable over the PTA observational period.
We now consider two representative GCs, M15 and $\omega$ Cen.

M15 is a core-collapsed GC located $10\,\mathrm{kpc}$ from Earth,
with a core radius of $r_c=0.41\,\mathrm{pc}$. 
The existence of an IMBH in M15 remains inconclusive given current observations, 
with models estimating a possible mass ranging from $500\,M_{\odot}$ to $4000\,M_{\odot}$ \cite{2025NSRev..12..347H,2002AJ....124.3270G,2003ApJ...582L..21B}. So far,
15 pulsars have been discovered in M15, and $9$ of them are MSPs. Most pulsars are within a few core radii of M15, and some have projected distances as small as $0.05\,r_c$. 
The recent timing of them has reached a precision of about $20\,\mathrm{\mu s}$ \cite{2024ApJ...974L..23W,2025RAA....25g1001D}.

We assume that the IMBH binary in M15 has a total mass of
$m_1+m_2=4000\,M_{\odot}$, and a mass ratio $q=1$.  Within the range where our
approximation holds, a longer orbital period of the binary $P$ leads to a
larger $R(t)$. However, considering observational constraints, we assume that
$P=5\,\mathrm{yr}$ .  And we assume that MSPs are
distributed within a region where $b\geq0.04\,r_c$ and $|z_e|\leq3\,r_c$.
Moreover, we keep the orientation angles $\theta=\pi/4$ and $\phi=\psi=0$. 
We include both circular and eccentric ($e=0.7$) orbits.

\begin{figure}
    \centering
    \includegraphics[width=1\linewidth]{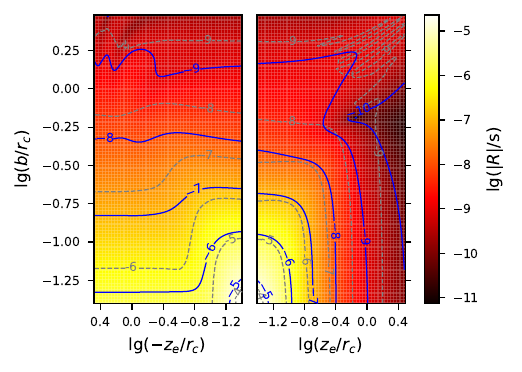}
    \caption{Timing residual for an MSP in M15 as a function of position, where $r_c=0.41\,\mathrm{pc}$ is the core radius of M15.
    We have assumed $m_1+m_2=4000\,M_{\odot}$, $q=1$, and $P=5\,\mathrm{yr}$.
    The dashed contours represent the timing residual for an elliptical orbit with $e=0.7$, with value taken as the maximum residual over one orbital period.
    }
    \label{fig:M15_5yr}
\end{figure}

The resulting magnitude of $R(t)$ as a function of the pulsar position is shown
in Fig.~\ref{fig:M15_5yr}, where the color-filled contour plot with blue solid contours depicts the circular orbit case. The gray dashed contours correspond to eccentric orbit case with $e=0.7$.
We find that, for circular orbits, $R(t)$ can exceed
$100\,\mathrm{ns}$ at $b\lesssim0.2\,r_c$, and reach $1\,\mathrm{\mu s}$ at
$b\lesssim0.1\,r_c$. Besides, $R(t)$ decays with increasing $b$, and
also drops rapidly for $z_e>0$. Therefore, we conclude that the MSPs projected 
near the GC center, well within the core radius, are likely to exhibit a detectable timing residual,
if the IMBH binary exists.
The existence of orbital eccentricity generally amplifies the magnitude of the residual. For those MSPs located within $0.3\,r_c$, the residual from an eccentric binary increases by about an order of magnitude.

$\omega$ Cen is the largest known GC in the Galaxy with a core radius of
$r_c=3.58\,\mathrm{pc}$. It is well-observed due to its relatively close
distance of $5.2\,\mathrm{kpc}$ from Earth. Therefore, $\omega$ Cen is one of
the best-studied GCs and is considered a prime candidate to host an IMBH in its
core \cite{2020ARA&A..58..257G}.  
Observations of fast-moving stars in the central $0.08\,\mathrm{pc}$ of
$\omega$ Cen imply a lower limit of $8.2\times10^3\,M_{\odot}$ to the IMBH mass
\cite{Haberle_2024}.  In contrast, studies on the motion and distribution of
MSPs point to an extended central mass and put the IMBH mass below
$6\times10^3\,M_{\odot}$ \cite{2025A&A...693A.104B}.  A total of 18 MSPs have
been discovered and confirmed in $\omega$ Cen, most of which are located within
one core radius. The smallest projected distance among them is $0.2\,r_c$
\cite{chen2023meerkat}. Observations of these MSPs have achieved a timing
precision of $\sim10\,\mathrm{\mu s}$ \cite{dai23}. 

\begin{figure}
    \centering
    \includegraphics[width=1\linewidth]{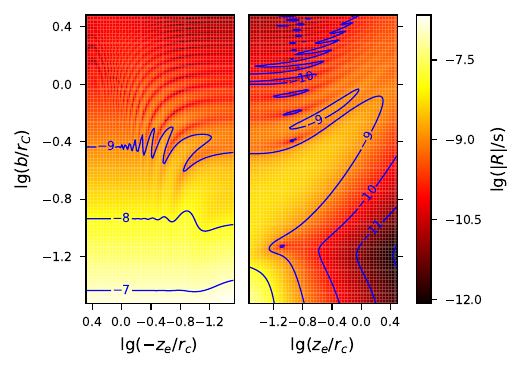}
    \caption{Timing residual for an MSP in $\omega$ Cen as a function of position, where $r_c=3.58\,\mathrm{pc}$ is the core radius of $\omega$ Cen.
    Here we use $m_1+m_2=10^4\,M_{\odot}$, $q=1$, and $P=5\,\mathrm{yr}$.
    }
    \label{fig:OmegaCen_5yr}
\end{figure}

Given the uncertainties above, we take $m_1+m_2=10^4\,M_{\odot}$ and
$q=1$. Similar to the M15 case, we assume an orbital period of $P=5\,\mathrm{yr}$
and an orientation of $\theta=\pi/4$. We further assume that MSPs are distributed in the
range where $b\geq0.03\,r_c$ and $|z_e|\leq 3\,r_c$. The result is shown in
Fig.~\ref{fig:OmegaCen_5yr}.  We see a somewhat weaker response than the M15 case
at the same $r_c$-normalized position, mainly due to the larger core radius. 
Given a distance of $0.2\,r_c$ from the center, the 
timing residual of an MSP is expected to reach $\sim 10\,\mathrm{ns}$, which is
much more challenging to detect than in the case of M15 MSPs.

\subsection{Timing Residuals Around Sgr A$^*$}

Now we study MSPs in the nuclei of galaxies. Sgr A$^*$, located at the center
of the Galaxy at a distance of $8\,\mathrm{kpc}$, is an SMBH with a mass of
$4\times10^6\,M_{\odot}$ \cite{gillessen2009monitoring,ghez2008measuring}.
Previous X-ray and $\gamma$-ray observations hint at a
population of $\sim10^3$ MSPs located within $1\,\mathrm{pc}$
\cite{wharton2012multiwavelength,bednarek2013gamma}.  The existence of MSPs in
the Galactic Center is corroborated by a theoretical cluster-inspiral model,
which predicts a population of $\sim2.7\times10^3$ MSPs within a radius of
$20\,\mathrm{pc}$.  However, according to this model, only $\sim5\%$ of the
MSPs lie in the central parsec.  Of these hundreds of MSPs, approximately $50$
are expected to be detectable by SKA, among which about $2$ are predicted to
lie within $1\,\mathrm{pc}$ \cite{macquart2015detecting,abbate2018probing}.
Timing these pulsars can help constrain the mass and distance of the possible
companion of Sgr A$^*$. 

\begin{figure}
    \centering
    \includegraphics[width=1\linewidth]{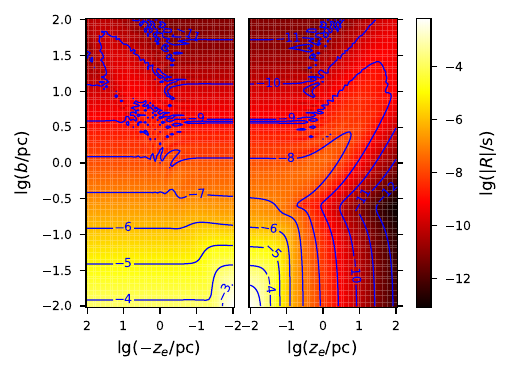}
    \caption{Timing residual for an MSP in Sgr A$^*$ as a function of position. 
    We have assumed $m_1+m_2=4\times10^6\,M_{\odot}$, $q=10^{-4}$, and $P=5\,\mathrm{yr}$.}
    \label{fig:SgrA_5yr_200_q00001}
\end{figure}

The observation of the S2 star over $23$ yrs has excluded an IMBH companion
with $m\gtrsim400\,M_{\odot}$ at $a\lesssim200\,\mathrm{au}$
\cite{will2023constraining}.  Therefore, for our calculation, we assume that
the binary has a total mass of $m_1+m_2=4\times10^6\,M_{\odot}$, a mass ratio
of $q=10^{-4}$, and an orbital period of $5\,\mathrm{yr}$.  The merger
timescale of the system is $T_{\mathrm{GW}}\sim2.34\times10^{12}\,\mathrm{yr}$.
The other parameters are set identically to those in M15. Given the above
estimation of the MSPs' distribution, we consider MSPs located at distances
ranging from $0.01\,\mathrm{pc}$ to $100\,\mathrm{pc}$ from the binary. The
result is shown in Fig.~\ref{fig:SgrA_5yr_200_q00001}. We find that the MSPs within
$1\,\mathrm{pc}$ have $R(t)\gtrsim10\,\mathrm{ns}$, and $R(t)$ can exceed
$1\,\mathrm{\mu s}$ for the MSPs within $0.1\,\mathrm{pc}$. 
At such small distances, MSPs may be accelerating due to
the gravitational pull of Sgr A$^*$. The consequence for pulsar timing
will be discussed in Section~\ref{sec:dis}.

\subsection{Timing Residuals in M31}

M31 is the nearest massive spiral galaxy, located at a distance of
$765\,\mathrm{kpc}$ from Earth. Observations indicate that M31 hosts an
SMBH in the center with a mass within the range of
$(0.5\sim1.4)\times10^8\,M_{\odot}$ \cite{bender2005hst,menezes2012discovery}.
Although the current observations have not resolved any individual pulsar in M31,
simulations suggest that telescopes such as FAST and SKA have the potential
to detect MSPs there \cite{smits2009pulsar}.

\begin{figure}
    \centering
    \includegraphics[width=1\linewidth]{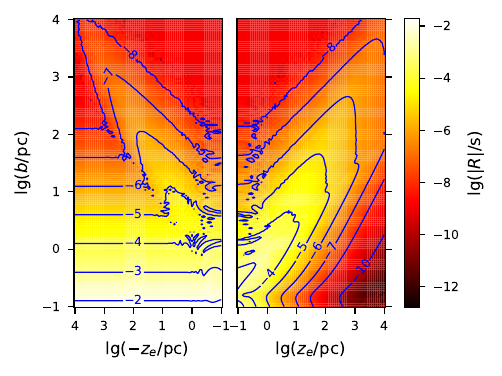}
    \caption{Timing residual for an MSP in M31 as a function of position. 
    We have assumed $m_1+m_2=10^8\,M_{\odot}$, $q=0.01$, and $P=3\,\mathrm{yr}$.
    }
    \label{fig:M31_3yr_200_q001}
\end{figure}

We assume that the SMBH binary in M31 has a total mass of
$m_1+m_2=10^8\,M_{\odot}$, a mass ratio of $q=0.01$, and an orbital period of
$P=3\,\mathrm{yr}$. Other parameters are set identically to those in M15.  The
merger timescale of the system is
$T_{\mathrm{GW}}\sim2.86\times10^7\,\mathrm{yr}$.  We consider MSPs located at
distances ranging from $0.1\,\mathrm{pc}$ to $10\,\mathrm{kpc}$ from the center
of M31.  The timing residuals are shown in Fig.~\ref{fig:M31_3yr_200_q001}.
Due to the binary's large mass, the emitted GWs are strong enough to produce
significant timing residuals ($\gtrsim 1\,\mathrm{\mu s}$) even for an MSP
located at $100\,\mathrm{pc}$ from the center.  Moreover, $R(t)$ can reach
$\gtrsim1\,\mathrm{ms}$ for an MSP with $b\lesssim0.5\,\mathrm{pc}$.  The
magnitudes of these timing residuals qualitatively differ from those derived in
the absence of non-radiative self-fields or pulsar terms
\cite{2025ApJ...978..104G}.  Our results suggest that finding MSPs in M31 and
timing them can help us effectively constrain the binarity of the central SMBH.

\subsection{Phase Difference Between MSPs}

PTAs detect GWs by revealing correlated patterns in the timing residuals of different pulsars. 
Therefore, not only are the magnitudes of the timing
residuals at different positions important, but also their phase differences.

\begin{figure}
    \centering
    \includegraphics[width=1\linewidth]{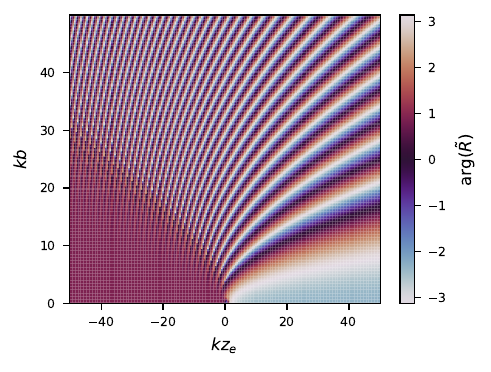}
    \caption{Relative phase of the timing residual as a function of position. We assume a circular GW source with $\theta=\phi=\psi=0$.}
    \label{fig:Phase_200}
\end{figure}

Fig.~\ref{fig:Phase_200} shows how the relative phase depends on pulsar positions.
The bottom-left and bottom-right corners of the plot show small variation. 
The cause can be found in Fig.~\ref{fig:Rer12}, which shows that the GW effect is concentrated around $kz\approx0$ for small $b$. Therefore, outside this central region, the phase and magnitude of the 
timing residual are largely insensitive to the value of $z_e$.
The rest part of the plot shows regular, stripe-like patterns. These stripes are more densely
spaced in the region $z_e<0$ than in $z_e>0$, 
because when $z_e<0$, the light is traveling in a direction opposite to the propagation direction of the GW so that 
the integrand
$\tilde{r}_{1,n=2}$ is more oscillatory.

\section{Discussion and Conclusion}\label{sec:dis}

Motivated by (i) the abundance of MSPs in GCs, (ii) the observational prospect
of discovering MSPs in the centers of the Milky Way and M31, and (iii) the
theoretical expectation that binary GW sources reside in these dense stellar systems, we
investigate the response of an array of MSPs to a GW source inside the PTA. Our
work completes the earlier ones by integrating four ingredients in the same
model: (i) eccentricity of the binary, (ii) second-order stress-energy tensor,
(iii) non-radiative self-field of the binary, and (iv) the pulsar term. They
are essential for modeling a realistic system, especially when the pulsar is
only a few wavelengths away from the GW source.

\begin{figure}
    \centering
    \includegraphics[width=1\linewidth]{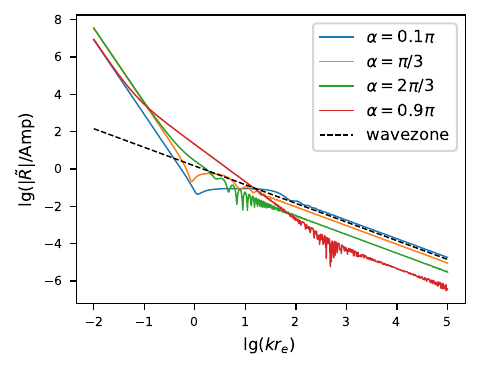}
    \caption{Dependence of $R/\mathrm{Amp}$ on $kr_e$ in our model (solid curves) and in the model
which neglects the pulsar term and the self-field (dashed line). 
Here $\alpha$ denotes the angle between $\mathbf{r_o}$ and $\mathbf{r_e}$.
}
    \label{fig:comparison}
\end{figure}

Recent works sometimes neglect the pulsar term and the self-field close
to the GW source \cite{2017PhRvD..96l3016M,2022ApJ...939...55G}.
In this case, the timing residual is affected only by the far-field wave-like component
during the propagation of the pulse. By averaging over the
pulsar direction on a sphere of radius $r_e$ (centered on the source) and over
source orientation, we find that the residual is proportional to $1/r_e$, or more specifically 
\begin{align}
    \langle R\rangle\sim\bar{\chi}\frac{h_o}{\omega}=0.365\times4\,\frac{\mathrm{Amp}}{kr_e},
\end{align}
where $\bar{\chi}=0.365\,r/r_e$ is a factor that represents the geometric effect of the relative position of the MSP and the source when $r\gg r_e$,
and $h_o$ is the characteristic strain at the observer \cite{2022ApJ...939...55G}.
This result, derived from a simplified model,
is more or less consistent with our model prediction when $kr_e\gg10$, as is shown in 
Fig.~\ref{fig:comparison} where $\alpha$ represents the angle between $\mathbf{r_o}$ and $\mathbf{r_e}$.
However, when $kr_e<10$, our full model predicts a much steeper
rise (between $1/r_e^2$ and $1/r_e^4$) of timing residual than $1/r_e$, highlighting the importance of the pulsar term
and the self-field. 
The $1/r^4$ term becomes insignificant around $kr\sim 1$. 
The $1/r^2$ term, however, can remain significant out to much larger $kr$, depending on the MSP's relative position.
In particular, when the MSP lies behind the GW source ($\alpha\sim\pi$) with a relatively small $kb$, it can remain significant up to $kr\sim10^2$ or even near $10^3$.
This is because, according to Appendix~\ref{app:R1(t)}, the $1/r^2$ contribution is of order $1/(kb)^2$ instead of $1/(kr)^2$, and the coefficient of the $1/r^2$ term is larger relative to the $1/r$ terms especially when $\alpha\sim\pi$.

The dependence of timing residual $R(t)$ on $M$ and $q$ is given directly by
Eq.~(\ref{eq:Amp}).  However, it is worth noting that although the binary's
period $P$ appears in Eq.~(\ref{eq:Amp}), the dependence of $R(t)$ on $P$ is
not fully captured by $\mathrm{Amp}$ alone.  Besides the scaling amplitude, the
dimensionless distance $kr=2\pi r/cP$ which appears in other terms is also
$P$-dependent.  Therefore, the proper scaling of $P$ must also include the
dependence of $R(t)$ on $kr$ shown in Fig.~\ref{fig:comparison}.  The
dependence evolves from $R(t)\propto P^{1/3}$ in the far field ($1/r$ zone) to $P^{4/3}$ in the
$1/r^2$ zone and $P^{10/3}$ in the $1/r^4$ zone.  

As shown in Fig.~\ref{fig:M15_5yr}, orbital eccentricity generally
amplifies the timing residual. The reason is the following. For highly
eccentric orbits ($e\gtrsim0.7$), the $1\omega$ harmonic dominates over the
$2\omega$ harmonic, so the binary can be approximated as a source with period
$2P$. According to the scaling relation derived above, the timing residual is
amplified by a factor of $\sim1.2$ in the $1/r$ zone, $\sim2.5$ in the $1/r^2$
zone, and $\sim10$ in the $1/r^4$ zone.

Applying our model to GC systems such as M15 and $\omega$ Cen, as well as the
nuclei of the Milky Way and the nearby galaxy M31, we find that for MSPs
located within the near field regime, the timing residual can reach
$100\,\mathrm{ns}$ to $\mathrm{\mu s}$ in GCs, and $\mathrm{\mu s}$ to
$\mathrm{ms}$ in galactic centers. The exact value depends on source mass and
scaled distance. For comparison, the timing residuals induced by the same GW
sources on the MSPs in the Galactic field (conventional PTA) fall in the range of
$10^{-3}-10$ ns because the characteristic strain is of the order
$10^{-20}\sim10^{-16}$. Detecting such signals is challenging using
conventional PTA method. This comparison highlights the advantage of using a miniature
PTA in a star cluster or galactic nucleus to detect potential GW sources inside the same
system. 

According to our previous work \cite{chen25}, detecting a binary GW source in a
star cluster with false alarms less than $5\%$ requires that the amplitude of
the timing residual should reach $A\gtrsim4.65\sigma/\sqrt{NN_p}$, where $N_p$
is the number of MSPs in the near field (in the same stellar cluster), $N$
denotes the total number of measurements, and $\sigma$ is the precision of each
measurement.  If we assume $10$ MSPs in the cluster and a total of $100$ 
measurements over an observational period of five years, 
a measurement precision of about
$\sigma\lesssim6.8\,\mathrm{\mu s}$ will enable us to 
detect a signal of
$A=1\,\mathrm{\mu s}$. Interestingly, such a precision is
becoming achievable in recent years in timing observations of MSPs in GCs
(e.g., \cite{wang2020discovery,2021ApJ...915L..28P,2025ApJS..279...51L}). A longer observational period will further
enhance the detectability by increasing $N$.

In our calculation we have assumed that the MSPs are not moving relative to the GW source.
However, those close to the source may undergo Keplerian motion and show additional timing residuals. 
For such an MSP, due to the gravity of the nearby GW source, the
orbital period $P_{\mathrm{orb}}$ around the GW source is given by
\begin{align}
    P_{\mathrm{orb}}\simeq&9.4\times10^2\,\mathrm{yr}\,\left(\frac{r_{e}}{0.01\,\mathrm{pc}}\right)^{3/2}\left(\frac{M}{10^4\,M_{\odot}}\right)^{-1/2}\\
    =&9.4\times10^3\,\mathrm{yr}\,\left(\frac{r_{e}}{1\,\mathrm{pc}}\right)^{3/2}\left(\frac{M}{10^8\,M_{\odot}}\right)^{-1/2},
\end{align}
where $r_e$ denotes the average distance from the MSP to the GW source.
According to the above estimations, an MSP over a typical observational period
of five years has approximately a constant velocity $v$, corresponding to a constant
Doppler shift $\Delta\nu$.  This induces a linear, non-periodic accumulation in
$R(t)$, which can be removed from the timing model.  

In the future, as the sensitivity of radio telescopes improves, a
microsecond-level timing residual may become detectable for the MSPs in GCs and
galactic nuclei.  Then the model developed here will enable direct constraints
on the binarity of the intermediate-mass and supermassive black holes in these
systems, revealing GW sources that would otherwise remain invisible to
traditional PTA analyses.

\begin{acknowledgments}
This work is supported by the National Key Research and Development Program of
China (Grant No. 2024YFC2207300) and the National Natural Science Foundation of
China (Grant No. 12473037). XC thanks Ver\'{o}nica V\'{a}zquez-Aceves, Siyuan
Chen, Kejia Lee, and Yanjun Guo for earlier discussions. HQ thanks Pau
Amaro-Seoane for comments on the timing-residual equaitons.
\end{acknowledgments}

\appendix

\section{Expressions of orientation factors}\label{app:qijn}

The rotation matrices are given by
\begin{align}
    &R_3(\psi)=
    \begin{pmatrix}
        \cos\psi&-\sin\psi&0\\
        \sin\psi&\cos\psi&0\\
        0&0&1
    \end{pmatrix},\\
    &R_2(\theta)=
    \begin{pmatrix}
        1&0&0\\
        0&\cos\theta&-\sin\theta\\
        0&\sin\theta&\cos\theta
    \end{pmatrix},\\
   &R_1(\phi)=
    \begin{pmatrix}
         \cos\phi&-\sin\phi&0\\
        \sin\phi&\cos\phi&0\\
        0&0&1
    \end{pmatrix}.
\end{align}

For a Keplerian motion in the $x-y$ plane, $\mathbf{x}'\mathbf{x}'^{T}$ can be written in the form of a Fourier series:
\begin{align}
    \mathbf{x'}\mathbf{x'}^{T}&=
    \begin{pmatrix}
        x'^2&x'y'&0\\
        x'y'&y'^2&0\\
        0&0&0
    \end{pmatrix}=\Real(\sum_{n=0}^\infty a^2\tilde{A}_n\mathrm{e}^{\mathrm{i}n\omega t}).
\end{align}

In our case, the nonzero components of $\tilde{A}_n$ are given by \cite{maggiore07}
\begin{align}
    \tilde{A}_{xx,n}&=\frac{1}{n}[J_{n-2}(ne)-J_{n+2}(ne)\nonumber\\&\qquad\qquad\quad-2e(J_{n-1}(ne)-J_{n+1}(ne))],\\
    \tilde{A}_{yy,n}&=\frac{1-e^2}{n}[J_{n+2}(ne)-J_{n-2}(ne)],\\
    \tilde{A}_{xy,n}&=\tilde{A}_{yx,n}=-\mathrm{i}\frac{\sqrt{1-e^2}}{n}[J_{n+2}(ne)+J_{n-2}(ne)\nonumber\\&\qquad\qquad\quad-e(J_{n+1}(ne)+J_{n-1}(ne))],
\end{align}
where $J_n$ is the $n$-th order Bessel function of the first kind.

\section{Analytical form of $\tilde{R}(t)$}\label{app:R1(t)}

Here we show the analytical expression of $\tilde{R}(t)$ from Eq.~(\ref{eq:R1t}, \ref{eq:R2t}, \ref{eq:xin}, \ref{eq:zetan}, \ref{eq:etan}). The complex amplitude $\tilde{R}_{1,n}$ is given by
\begin{align}
    \tilde{R}_{1,n}=\int_{kz_e}^{kz_o}\,\mathrm{d}(kz)\frac{\tilde{\xi}_n(kz)}{(kr)^3}\mathrm{e}^{\mathrm{i}n(kz-kr-kz_o)}\label{eq:R1n}.
\end{align}

To integrate analytically, it is more convenient to introduce a dimensionless variable $x$ and a scale factor $\beta$, defined by
\begin{align}
    x=&\frac{\omega}{c}(\sqrt{z^2+b^2}-z)=kr-kz,\\
    \beta=&\frac{\omega b}{c}=kb=\left.\frac{v}{c}\right/\frac{a}{b},
\end{align}
and Eq.~(\ref{eq:R1n}) becomes
\begin{align}
    \tilde{R}_{1,n}=\int_{x_o}^{x_e}\frac{4x\tilde{\xi}_n(x)}{(x^2+\beta^2)^2}\exp[-\mathrm{i}n(x+kz_o)]\,\mathrm{d}x.
\end{align}

We can rewrite $\tilde{R}_{1,n}$ by separating the orientation factor $\tilde{q}_{ij,n}$:
\begin{align}
    \tilde{R}_{1,n}=\sum_{ij}[\tilde{U}_{ij,n}(x_e)-\tilde{U}_{ij,n}(x_o)]\tilde{q}_{ij,n},
\end{align}
and the nonzero components of $\tilde{U}_{ij,n}(x)$ are
\begin{widetext}
    \begin{align}
        \tilde{U}_{xx,n}(x)&=\frac{2\exp[-\mathrm{i}n(x+kz_o)]}{(x^2+\beta^2)^3}[(x^2+\beta^2)^2-2\mathrm{i}\beta^2(-3\mathrm{i}x^2+nx^3-\mathrm{i}\beta^2+nx\beta^2)]\\
        &=\left[\frac{(kz)^2}{(kr-kz)(kr)^3}-\frac{(kb)^2}{(kr-kz)^2(kr)^2}-\mathrm{i}n\frac{(kb)^2}{(kr-kz)(kr)^2}\right]\mathrm{e}^{\mathrm{i}n(kz-kr-kz_o)}\\
        &=\left[\frac{(kz)^2(kr+kz)}{(kb)^2(kr)^3}-\frac{(kr+kz)^2}{(kb)^2(kr)^2}-\mathrm{i}n\frac{kr+kz}{(kr)^2}\right]\mathrm{e}^{\mathrm{i}n(kz-kr-kz_o)},\\
        \tilde{U}_{yy,n}(x)&=\frac{2\exp[-\mathrm{i}n(x+kz_o)]}{(x^2+\beta^2)}\\
        &=\frac{1}{(kr-kz)kr}\mathrm{e}^{\mathrm{i}n(kz-kr-kz_o)}=\frac{kr+kz}{(kb)^2kr}\mathrm{e}^{\mathrm{i}n(kz-kr-kz_o)},\\
        \tilde{U}_{zz,n}(x)&=\frac{4\exp[-\mathrm{i}n(x+kz_o)]}{(x^2+\beta^2)^3}x^2(-x^2-\mathrm{i}nx^3+\beta^2-\mathrm{i}nx\beta^2)\\
        &=\left[\frac{kz}{(kr)^3}-\mathrm{i}n\frac{kr-kz}{(kr)^2}\right]\mathrm{e}^{\mathrm{i}n(kz-kr-kz_o)}\\
        &=\left[\frac{kz}{(kr)^3}-\mathrm{i}n\frac{(kb)^2}{(kr+kz)(kr)^2}\right]\mathrm{e}^{\mathrm{i}n(kz-kr-kz_o)},\\
        \tilde{U}_{zx,n}(x)&=\tilde{U}_{xz,n}(x)=\frac{4\exp[-\mathrm{i}n(x+kz_o)]}{(x^2+\beta^2)^3}\beta x^2(2x+\mathrm{i}nx^2+\mathrm{i}n\beta^2)\\
        &=\left[\frac{kb}{(kr)^3}+\mathrm{i}n\frac{kb}{(kr)^2}\right]\mathrm{e}^{\mathrm{i}n(kz-kr-kz_o)},
    \end{align}
\end{widetext}
which is equivalent to the results in \cite{jenet05}.

Similarly, for $\tilde{R}_{2,n}$ we have
\begin{align}
    \tilde{R}_{2,n}&=-\frac{\mathrm{e}^{\mathrm{i}n(kz-kz_o-kr)}}{2n^2(kr)^3}\biggl[\biggl(\mathrm{i}n \biggl(\frac{kz}{kr}+1\biggr)\nonumber\\
    &\qquad\qquad+3\frac{kz}{(kr)^2}\biggr)\tilde{\zeta}_n
    -\partial_{kz}\tilde{\zeta}_n+2\mathrm{i}n\tilde{\eta}_{n}
    \biggr]\Biggr|^{e}_{o}\\
    &=\sum_{i,j}[\tilde{V}_{ij,n}(z_e)-\tilde{V}_{ij,n}(z_o)]\tilde{q}_{ij,n},
\end{align}
and the nonzero components of $\tilde{V}_{ij,n}(z)$ are
\begin{widetext}
    \begin{align}
        \tilde{V}_{xx,n}=&-\frac{\mathrm{e}^{\mathrm{i}n(kz-kz_o-kr)}}{2n^2(kr)^3}\biggl[15\frac{(kb)^2kz}{(kr)^4}-3\frac{kz}{(kr)^2}+\mathrm{i}n\biggl(15\frac{(kb)^2kz}{(kr)^3}+3\frac{(kb)^2}{(kr)^2}-3\frac{kz}{kr}-1\biggr)\nonumber\\
        &+n^2\biggl(-6\frac{(kb)^2kz}{(kr)^2}-3\frac{(kb)^2}{kr}+kr\biggr)+\mathrm{i}n^3\biggl(-\frac{(kb)^2kz}{kr}-(kb)^2-(kz)(kr)-(kr)^2\biggr)\biggr],\\
        \tilde{V}_{yy,n}=&-\frac{\mathrm{e}^{\mathrm{i}n(kz-kz_o-kr)}}{2n^2(kr)^3}\biggl[-3\frac{kz}{(kr)^2}+\mathrm{i}n\biggl(-3\frac{kz}{kr}-1\biggr)+n^2(kr)-\mathrm{i}n^3((kr^2)+(kz)(kr))\biggr],\\
        \tilde{V}_{zz,n}=&-\frac{\mathrm{e}^{\mathrm{i}n(kz-kz_o-kr)}}{2n^2(kr)^3}\biggl[15\frac{(kz)^3}{(kr)^4}-9\frac{kz}{(kr)^2}+\mathrm{i}n\biggl(15\frac{(kz)^3}{(kr)^3}+3\frac{(kz)^2}{(kr)^2}-9\frac{kz}{kr}-1\biggr)\nonumber\\
        +&n^2\biggl(-6\frac{(kz)^3}{(kr)^2}-3\frac{(kz)^2}{kr}+6kz+kr\biggr)+\mathrm{i}n^3\biggl(-\frac{(kz)^3}{kr}-(kz)^2+3(kz)(kr)-(kr)^2\biggr)\biggr],\\
        \tilde{V}_{zx,n}+&\tilde{V}_{xz,n}=-\frac{\mathrm{e}^{\mathrm{i}n(kz-kz_o-kr)}}{2n^2(kr)^3}\biggl[30\frac{(kz)^2kb}{(kr)^4}-6\frac{kb}{(kr)^2}+\mathrm{i}n\biggl(30\frac{(kz)^2kb}{(kr)^3}+\frac{6(kz)(kb)}{(kr)^2}-6\frac{kb}{kr}\biggr)\nonumber\\
        &+n^2\biggl(-12\frac{(kz)^2kb}{(kr)^2}-6\frac{(kz)(kb)}{kr}+6kb\biggr)+\mathrm{i}n^3\biggl(-2\frac{(kz)^2kb}{kr}+4(kb)(kr)-2(kb)(kz)\biggr)\biggr].
    \end{align}
\end{widetext}
This result differs from Eq.~(8) in \cite{jenet05} by an additional term $(\frac{3x_ix_j}{2r^2}+\delta_{ij})\dot{Q}_{ij}/r^2$.

\bibliography{ptabib}
\end{document}